\documentstyle[preprint,eqsecnum,aps]{revtex}

\def\kbar{\rlap\slash k}

\begin{document}
\draft
\title{Generalized Relativistic Meson Wave
Function}
\author{A. Szczepaniak, Chueng-Ryong Ji and Stephen.~R.~Cotanch}
\address{Department of Physics, North Carolina State University,
Raleigh, North Carolina 27695-8202} \date{\today} \maketitle
\begin{abstract}
We study the most general, relativistic, constituent $q{\overline q}$
meson wave function within a new covariant
 framework. We find that by including a tensor wave function component, a
pure valence quark model
is now capable of reproducing not only all static pion data ($f_\pi$,
$\langle r_\pi^2 \rangle$) but also the
distribution amplitude, form factor $(F_\pi(Q^2))$,
and structure functions. Further, our generalized spin wave function
provides a much better detailed description of meson properties
than
 models using a simple relativistic extension of the
$S=L=0$ nonrelativistic
 wave function.
\end{abstract}
\narrowtext

\section{INTRODUCTION}
\label{sec:Intro.}

Since an exact solution to a bound state problem in QCD is still
unavailable many approximate treatments have been developed. Among
them, the constituent quark model has perhaps received the most attention
and is widely regarded as a very
efficient and effective tool in description of hadronic
phenomena~\cite{qm1,qm2,qm3}. The
complex low energy structure of QCD currently
precludes an
unambiguous identification of the complete degrees of freedom
and it is therefore important to continue to advance, refine and test the
 valence quark dominance approximation.
Prospects are encouraging that new,
precision data provided by future CEBAF experiments will
significantly clarify this situation and also detail the role of
exotic quark and/or gluon configurations. However,
before the relative importance
of valence versus exotic configurations can be established,
uncertainties in the description of hadronic amplitudes due to
valence quark model approximations must be reduced.
In a previous study ~\cite{AS} we investigated alternative model
approaches by comparing
different relativistic formulations
for the light pseudoscalar mesons. In particular, we developed and
examined  a covariant variable front approach which permitted quantitatively
assessing the relevance of Lorentz covariance for any constituent
formulation.
We also established that
 pion and the kaon static properties can be well
described by relativistic
 models utilizing
 constituent $q{\overline q}$ meson wave functions
 represented by the product of a noninteracting spinor component
and a momentum space orbital amplitude.
It is believed that such models will be able to
describe low energy mesonic data to within a 10-20\%.
 Unfortunately, there is still a large energy gap
 between this region and the energy-momentum scale where
asymptotic freedom dominates~\cite{pqcd} and
 a clear need exists for an improved, QCD related quark model to describe
this intermediated regime. This is especially true for analyzing
 electromagnetic form factors with $Q^2 > 2-3 \mbox{ GeV}^2$ for
which all quark models described in Ref.~\cite{AS} fail to
describe the data. As mentioned above,
however, before making any exotic extensions of the quark model an improved
 framework for the $q{\overline q}$ system must be developed.
The purpose of the present paper is to report one such attempt which
utilizes a more
 general quark meson wave functions.

In this article we extend our covariant variable front quark model
by incorporating a more general spinor
wave function with tensor components. Although not an observable,
the wave function is
 constrained by results
 from QCD studies of moments
of the distribution amplitudes~\cite{CZ,RAD,LATT1,LATT2}.
Additional information comes from meson structure functions
measured in Drell-Yan experiments~\cite{DY1,DY2,Hadr}.
In the next section we briefly
review the basic assumptions of the covariant quark model, and detail our
extended valence meson wave function. In Sec. III we
analyze the quark distribution amplitudes and structure functions for the
low lying mesons and present numerical results. Finally, we discuss and
summarize our major results in Sec.IV

\section{VALENCE QUARK MESON WAVE FUNCTION}

Following our previous paper we specify the quantization surface $\Sigma$
by a timelike four vector $n^\mu$
with $n^2 = 1$, although the analysis is also appropriate for
 the case when $n^\mu$ is
null-like, $n^2=0$.
The transverse and longitudinal components of an arbitrary four vector
$A^\mu$ are  denoted as $A_T$ and $A_L$, respectively, $A^\mu = (A_L,A_T)$
and are defined by

\[
 A_L \equiv n\cdot A\;, A^\mu_T \equiv A^\mu - A_L n^\mu.
\]
Since $n\cdot A_T=0$, $A^\mu_T$ has only three independent
components which we label $(A^\perp_T,A^3_T)$ and henceforth identify by
$A_T$.
We define the
wave function $\Psi^\alpha(k_{Ti};\lambda_i;\tau_i), i=1\dots N$
as the probability amplitude for finding $N$ constituents
(quarks, antiquarks, gluons) with transverse momenta $k_{Ti}$,
helicities $\lambda_i$ and flavor-color components $\tau_i$ in a
meson state $\alpha$, ($\alpha=1\dots 8$ for the pseudoscalar octet)
with momentum $P_T = \sum_i k_{Ti}, P_L = \sqrt{M^2 + |P^2_T|}$
 by the following matrix element

\begin{equation}
\langle k_{Ti};\lambda_i; \tau_i| P_T;\alpha \rangle
 =  (2\pi)^3\sqrt{2P_L}\delta^3(P_T - \sum_i k_{Ti})
\left[\prod_i\sqrt{{k_{iL}}\over {m_i}}\right]
\Psi^\alpha(k_{Ti};\lambda_i;\tau_i). \label{mvf}
\end{equation}
Here the single particle states $|k_T;\lambda;\tau \rangle$ describe
effective, massive constituents quantized on the spacelike
surface $\Sigma$ perpendicular to $n^\mu$ and the longitudinal
momenta are constrained by the on shell condition,

\begin{equation}
k_{iL} = \sqrt{m^2_i + |k^2_{iT}|}. \label{onshell} 
\end{equation}
The single particle states are normalized according to

\begin{equation}
\langle k'_T,\lambda',\tau'| k_T,\lambda,\tau \rangle = 
{{k_L}\over m} (2\pi)^3 \delta^3(k'_T - k_T)\delta_{\lambda'\lambda}
\delta_{\tau'\tau} \label{fnor}
\end{equation}
for fermions and 

\begin{equation}
\langle P'_T, \alpha| P_T, \beta \rangle = 2 (2\pi)^3 P_L \delta^3(P'_T -
P_T) \delta_{\alpha\beta} \label{snor}
\end{equation}
for bosons.

The set of wave functions defined by Eq.~(\ref{mvf}) constitutes a
 Lorentz group representation basis. In a formalism with a fixed
quantization surface,
 Lorentz transformations depending on interactions do not leave
$\Sigma$ invariant
interactions
do not conserve particle number and therefore as the system evolves
different Fock
 sectors mix~\cite{BRHUANG}. This in general leads to complicated
transformation properties for the wave functions under the action of
interaction dependent
generators of the Lorentz group. A simplification is usually made
by using an interaction free transformation rule
for the wave function in a given Fock sector~\cite{DZIEM}.
The valence $q{\overline q}$ wave functions describing a meson state
with 4-momenta
 $P$ and $P' = \Lambda P
 \equiv {\cal L}(P\to \Lambda P)P$ respectively are thus related by

\begin{equation}
{1\over \sqrt{2P_L}}\left[\prod_i\sqrt{{k_{iL}}\over
{m_i}}\right] \Psi^\alpha(k_{Ti};\lambda_i;\tau_i)
 =  \prod_i\sum_{\lambda'_i}D_{\lambda_i\lambda'_i}(R_W)
{1\over \sqrt{2P'_L}}\left[\prod_i\sqrt{{k'_{iL}}\over
{m_i}}\right] \Psi^\alpha(k'_{Ti};\lambda'_i;\tau_i) \label{trans}
\end{equation}
with $k'_i = \Lambda k_i$ and $D_{\lambda \lambda'}(R_W)$ denoting the
matrix representations of the Wigner rotation $R_W$ in the spin space
expressed in terms of $q{\overline q}$ momentum variables,

\[
R_W = R_W(k_{iT}) = {\cal L}(\Lambda P \to {\hat P})
{\cal L}(P \to \Lambda P){\cal L}(\hat{P} \to P),
\]
and ${\hat P} = (M,{\bf 0})$ being the meson rest frame momentum.
Using this approximation
  the meson state is
defined in a standard way and contains only
$q{\overline q}$ valence component. Eqs.~(\ref{mvf}),~(\ref{fnor}) and
then lead to the following wave function normalization

\begin{eqnarray}
& & \sum_{\lambda,\tau} \int [dk_{iT}]^{q{\overline q}}_P \Psi^{\dag
\alpha}(k_{iT};\lambda;\tau) \Psi^{\beta}(k_{iT};\lambda;\tau) =
\delta^{\alpha\beta}, \nonumber \\
& & [dk_{iT}]^{q{\overline q}}_P= [dk_{iT}]_P \equiv
\prod_{i}^2{{d^3k_{iT}}\over {(2\pi)^3}} 
 (2\pi)^3 \delta^3(P_T - \sum_{i}^2 k_{iT}). \label{norm}
\end{eqnarray} 
In general, approximations and specifically Fock space truncations destroy
 covariance.
Here the truncation generates noncovariance by the emergence
 of an unphysical dependence of matrix elements upon $n^\mu$.
In ~\cite{AS} we describe a method for restoring
 covariance by allowing the quantization surface $\Sigma$,
or equivalently the quantization vector $n^\mu$, to transform actively
under Lorentz transformations by relating $n^\mu$ to the meson external
momenta.

 The normalization Eq.~(\ref{norm}) is identical to
the corresponding nonrelativistic expression in the meson rest frame,
since the relativistic wave function is constructed
to reduce to the nonrelativistic one in this frame.
 For the ground sate pseudoscalar octet the form of the relativistic
wave function is then derived from Eq.~(\ref{trans}) to be

\begin{eqnarray}
& & \Psi^\alpha(k_{Ti};\lambda_i;\tau_i) =
\chi^\alpha_{\tau_1\tau_2}\xi(k_{iT};\lambda_i)
 \Phi({\cal M}^2), \nonumber \\ 
& & \chi^\alpha_{\tau_1\tau_2} \equiv
 i\left[ {\lambda^\alpha \over
\sqrt{2}}\otimes {I \over \sqrt{3}}\right], \nonumber \\
& & \xi(k_{iT};\lambda_i)
\equiv \sqrt{2} {\sqrt{m_1m_2} \over \sqrt{{\cal M}^2 - (m_1-m_2)^2}}
{\overline u}(k_{1T};\lambda_1)\gamma_5 v(k_{2T};\lambda_2) 
\label{wfun}
\end{eqnarray}
where $\lambda^\alpha$ are the Gell-Mann $SU(3)$ flavor matrices,
$I$ is the identity matrix in the color space, $m_1,m_2$ are the
quark and antiquark constituent masses respectively, ${\cal M}$ is
the $q{\overline q}$ invariant mass,

\begin{equation}
{\cal M}(k_i) \equiv (k_1 + k_2)^2 =
(k_{1L}(k_{1T}) + k_{2L}(k_{2T}))^2 + (k_{1T} + k_{2T})^2,
\end{equation}
with $k_{iL}(k_T)$ given be Eq.~(\ref{onshell}) and
$\Phi({\cal
M})$ being the spin independent orbital wave function usually assumed
gaussian,
\begin{equation}
\Phi({\cal M}) = N\exp\left[- {{\cal M}^2 \over {8\beta^2}}\right].
\label{orb}
\end{equation}
The overall normalization constant $N$ is determined from
Eq.~(\ref{norm}).

As explained above the explicit form of the transverse variables will
depend on the choice for $n^\mu$ which in turn will be specified after
selecting which matrix element is to be calculated with the
wave function of Eq.~(\ref{wfun}).
The wave function specified by Eq.~(\ref{wfun}) has also been extensively
studied in ~\cite{AS} for fixed quantization schemes. Here we study
the extension of Eq.~(\ref{wfun}) to the most general
Dirac structure for the $q{\overline q}$ system. We write the
spinor component, $\xi(k_{iT};\lambda_i)$ of the wave function in
the general form, 

\begin{equation} 
\xi(k_{iT};\lambda) = \sum_p {\overline u}
(k_{T1},\lambda_1)\Gamma_p v(k_{T2},\lambda_2).  \label{wfgeneral}
\end{equation} 
with the Lorentz $4\times 4$ matrices $\Gamma_p$ represented by combinations
 of Dirac matrices and the constituent momenta,
$k_i$. Using the Dirac equations for the free $u$ and $v$
spinors it is easily shown that sum in Eq.~(\ref{wfgeneral}) reduces
to two terms involving either
$\gamma_5$ or $[\kbar_1,\kbar_2]\gamma_5$.
The most general wave function $\Psi^\alpha$ for a pseudoscalar meson can
thus be written in the form

\begin{eqnarray}
& & \Psi^\alpha(k_{iT};\lambda;\tau_i) = \chi^\alpha_{\tau_1\tau_2}
 \sqrt{2} {\sqrt{m_1m_2} \over \sqrt{{\cal M}^2 -
(m_1-m_2)^2}} \nonumber \\
& &  {\overline u} (k_{T1},\lambda_1)
\left[\gamma_5\Phi_P({\cal M}) +  [\kbar_1,\kbar_2]\gamma_5
\Phi_T({\cal M})\right]
 v(k_{T2},\lambda_2).
\label{wfnew}
\end{eqnarray}
We shall assume that the two spin independent wave functions
appearing in Eq.~(\ref{wfnew}) can be parameterized with a gaussian
shape of Eq.~(\ref{orb}) having the same momentum size parameter
$\beta$

\begin{equation}
\Phi_P({\cal M}) = \Phi({\cal M})\;, \Phi_T({\cal M}) = r_{PT}{1\over {{\cal
M}\beta}} \Phi({\cal M}). \label{rel}
\end{equation}
Since the tensor term in Eq.~(\ref{wfnew})
explicitly involves higher powers of the constituent longitudinal and
transverse variables whose ranges are related to ${\cal M}$
and $\beta$,
respectively, we have chosen the ${\cal M}\beta$ factor as the relative
normalization between the $\Phi_P$ and $\Phi_T$ terms in Eq.~(\ref{rel}).
The dimensionless, numerical coeficient
$r_{PT}$ will then be determined by fitting various properties of the pion.
In the $r_{PT} = 0$ limit the wave function of Eq.~(\ref{wfnew})
reduces to the one of Eq.~(\ref{wfun}) and corresponds to the
$S=L=0$ state in the meson rest frame. The tensor term
introduces an $S=L=1$ orbital component which mixes the $S=L=1$
lower with $S=L=0$ upper components of the Dirac
spinors to give an overall $J^{P} = 0^-$ state.

\section{Numerical Results}
\subsection{$\pi$ electromagnetic form factor}

Before detailing the quark distributions given
by the generalized wave functions including the additional tensor
component $\Phi_T$ we shall
present new results for the pion electromagnetic form factor.
The form factor calculation
for both
 noncovariant and covariant
variable front models
has been previously summarized ~\cite{AS}. For simplicity we
maintain
the same quark masses and oscillator size parameter $\beta$ used in
 our previous calculations  with $m_q = \beta =
250\mbox{ MeV}$ and vary $r_{PT}$ to optimize the
form factor description. It is significant to note that the
 tensor term drastically modifies the
 form factor behaviour especially in the high momentum transfer region $Q^2
> 2 \mbox{ GeV}^2$.
Further as shown in Fig.~1 the generalized model provides an
 agreement with the data in this region without altering the correct
 low $Q^2$ behaviour.
As shown below the
  importance of the tensor
term is further demonstrated in the analysis
of the distribution amplitude and the structure functions.

\subsection{Distribution amplitude}
The quark distribution amplitude for a meson $\alpha$,
$\phi^\alpha(\xi)$, can in principle be obtained from the
 moments $\langle \xi^n \rangle$,~\cite{CZ}
\begin{equation}
\langle \xi^n \rangle  = \int d\xi \xi^n \phi^\alpha(\xi) \label{mtop}
\end{equation}
which are formally defined by the matrix elements

\begin{equation}
\langle 0| {\overline \psi}(0)\gamma^\mu\gamma_5{\lambda^\beta \over 2}
\stackrel{\leftrightarrow}{\partial}_{\mu_1}\dots
\stackrel{\leftrightarrow}{\partial}_{\mu_n}\psi(0) | P;\alpha \rangle
= if_\alpha \langle \xi^n \rangle \delta_{\alpha\beta}P^\mu P_{\mu_1} \dots
P_{\mu_n} + \mbox{trace} \label{da}
\end{equation}
where the trace terms corresponding to higher twist operators have been
omited. Here $f_\alpha$ is
the meson decay constant
and the normalization is such that $\langle \xi^0 \rangle =
1$.
 The
expansion of the quark field operators, $\psi,{\overline \psi}$ in terms of
constituent $q{\overline q}$ creation and annihilation operators
Eqs.~(\ref{wfnew}),~(\ref{rel}) leads to

\widetext

\begin{eqnarray}
& & \langle \xi^n \rangle  = {\sqrt{6}\over {f_\alpha}} \int {{d^3q_T}\over
{(2\pi)^3}}{1\over \sqrt{M_\alpha}} {{m_1m_2}\over
\sqrt{k_{1L}k_{2L}{\cal M}^2 - (m_1-m_2)^2)}}
 \left({{2q_T^3}\over M_\alpha} + {{k_{1L} - k_{2L}}\over
M_\alpha}\right)^n \Phi({\cal M}) \nonumber \\
& & \left[{ {k_{1L}m_2 + k_{2L}m_1 }\over {m_1m_2}}
 - r_{PT}{{2 ( (k_{1L}+k_{2L})^2 - (m_1+m_2)^2)}\over {\beta(m_1+m_2)}}
\left(1 - {{(m_1-m_2)(k_{1L}m_2-k_{2L}m_1)}\over
{(k_{1L}+k_{2L})m_1m_2}}  \right)\right]\nonumber \\
\label{moments}
\end{eqnarray}

\narrowtext
with  $q_T = (q^3_T,q^\perp_T)$,
\begin{eqnarray}
& & k_{iL} = \sqrt{m_i^2 + |q_T^2|}, \nonumber \\
& & {\cal M} = k_{1L} + k_{2L}, \label{calm}
\end{eqnarray}
and $M_\alpha$ being the meson mass. The decay constant $f_\alpha$ is
calculated from Eq.~(\ref{moments}) using $\langle \xi^0 \rangle = 1$
(see also Sec.V in Ref.~\cite{AS}).
In QCD, it can be shown that the physical part of the distribution
amplitude $\phi^\alpha(\xi)$ is restricted for $-1<\xi<1$~\cite{CZ} and
for large
$n$ the moments in Eq.~(\ref{da}) behave as $\langle \xi^n \rangle \to
1/n^2$  implying that
$\phi^\alpha(\xi \to \pm 1) \to 0$. In order to reproduce this feature in
our model calculation we make the replacement

\begin{equation}
M_\alpha \to {\cal M} = k_{1L} + k_{2L} = \sqrt{m_1^2+|q_T^2|} +
\sqrt{m_2^2+ |q_T^2|}. \label{pres} \end{equation}
In Table~1 we list values for both $\pi$ and $K$ mesons decay
constants calculated from Eq.~(\ref{moments}) using
spin averaged meson
masses $M_\pi = 610 \mbox{ MeV}$ and $M_K = 790 \mbox{ MeV}$ and
compare with results using dynamical masses for $M_\alpha$
determined by Eq.~(\ref{pres}). Note that
the experimental values lie almost midway between the
 two methods. The sensitivity to the mass
prescription for
 normalized quantities like moments of the distribution
amplitude $\langle \xi^n \rangle$ or structure functions, which we
analyze in the following subsection, is even smaller.
Using Eq.~(\ref{pres}) in Eq.~(\ref{moments}) we make the following
change of variables

\begin{equation}
q^3_T \to \xi(q_T) \equiv  {{2q^3_T}\over {\cal M}} + {{ k_{1L} - k_{2L}
}\over {\cal M}}. \label{qtoxi}
\end{equation}
For fixed $q^\perp_T$ the variable $\xi(q_T)$ maps the domain of the
variable
$q^3_T$, $-\infty<q^3_T<\infty$ into the finite interval $-1<\xi(q_T)<1$.
The expression for the distribution amplitude can now be obtained from
Eqs.~(\ref{mtop}), ~(\ref{moments}), ~(\ref{pres}) and ~(\ref{qtoxi})
and is given by

\widetext

\begin{eqnarray}
& & \phi^\alpha(\xi) =
{\sqrt{6}\over {f_\alpha}} \int {{d^2q^\perp_T}\over
{(2\pi)^3}}{1\over \sqrt{{\cal M}}} J(q^\perp_T,\xi){{m_1m_2}\over
\sqrt{k_{1L}k_{2L}{\cal M}^2 - (m_1-m_2)^2)}}
\Phi({\cal M}) \nonumber \\
& & \left[{ {k_{1L}m_2 + k_{2L}m_1 }\over {m_1m_2}}
 - r_{PT}{{2 ( (k_{1L}+k_{2L})^2 - (m_1+m_2)^2)}\over {\beta(m_1+m_2)}}
\left(1 - {{(m_1-m_2)(k_{1L}m_2-k_{2L}m_1)}\over
{(k_{1L}+k_{2L})m_1m_2}}  \right)\right]\nonumber \\
  \label{distampl}
\end{eqnarray}
\narrowtext
with $k_{iL}$ and ${\cal M}$, now functions of $q^\perp_T,\xi$,
obtained from Eq.~(\ref{calm}) using Eq.~(\ref{qtoxi}) to
express $q^3_T = q^3_T(\xi)$. $J(q^\perp_T,\xi)$ is the Jacobian of the
transformation from $(q^3_T,q^\perp_T)$ to
$(\xi,q^\perp_T)$. The explicit forms of the functions,
$k_{iL}(q^\perp_T,\xi)$,  ${\cal M}(q^\perp_T,\xi)$ and $J(q^\perp_T,\xi)$
are given in the Appendix.

In Figs.~2 and 3 we plot the function $\phi(\xi)$ for $\pi$ and $K$ mesons
respectively. In Fig.~2 the solid line gives our prediction for the set
of parameters $m_q=\beta=250\mbox{ MeV}$, $r_{PT}=1.0$ which best describe
 the form factor shown in
Fig.~1. The dashed line is the prediction for the pion distribution
amplitude without the tensor term in the wave
function of Eq.~(\ref{wfnew}), i.e. with $r_{PT}=0$. The wide, camel like
shape of the
distribution amplitude obtained for $r_{PT}=1.0$, a value that optimizes
the form factor description, is similar to that provided by the QCD
sum rule approach~\cite{CZ}
for the matrix element of Eq.~(\ref{da}), however, the predictions for
 the lowest moments $\langle \xi^2 \rangle = 0.27$, $\langle \xi^4 \rangle
=0.13$, are smaller then
the ones of Chernyak and Zhitnitsky ($\langle \xi^2 \rangle_{CZ} = 0.40$,
$\langle \xi^4 \rangle_{CZ}
= 0.24$) yet closer to those obtained in Ref.~\cite{RAD} and
lattice calculations~\cite{LATT1,LATT2}.
 The asymmetry in the $K$ distribution amplitude in Fig.~3 is due to
the large $SU(3)$ breaking due
strange quark mass ($m_s = 480 \mbox{ MeV}$).
Notice the sensitivity of the pion distribution amplitude to
$r_{PT}$ which is shown in Fig.~4. The largest sensitivity is observed
for $r_{PT} \sim 0.5$ where the interference between the pseudoscalar
and tensor terms dominate.

\subsection{Structure functions}
Structure functions contain important hadronic information and are obtained
from inelastic processes.
Accordingly we wish to further test our approach by computing the meson
structure functions and comparing with available data
 usually extracted from Drell-Yan
lepton~\cite{DY1,DY2} or charged hadron production~\cite{Hadr} on nuclear
targets using mesonic
beams. The extraction from hadron production
experiments
is typically more complicated and somewhat model dependent due to
uncertainties in the
 hadronization mechanism. In this paper the experimental
 pion structure function was extracted from
the Drell-Yan
muons produced by $252\mbox{ GeV}$ pions on tungsten~\cite{DY2}.
The theoretical cross section for muon production
as a function of longitudinal momentum fraction $\xi_F$ of the muon
pair
is given by
\begin{eqnarray}
& & { {d^2\sigma} \over {dQ^2d\xi_F}} = K {{4\pi\alpha^2} \over
{9 Q^4}}{{ [f^\pi_v(x_1)G^N_v(x_2)
 + f^\pi_s(x_1)G^N_s(x_2)]} \over {(\xi_F^2 + 4Q^2/s)^{1/2} } }, \nonumber
\\ & & x_{1,2} = [\pm \xi_F + (\xi^2_F + 4 Q^2/s)^{1/2}]/2, \label{cross}
\end{eqnarray}
where $Q^2$ and $s$ are the mass of the muon pair and the square
of the c.m energy respectively, and $\alpha$ is the
electromagnetic fine structure constant. $f^\pi_{v(s)}(x_1)$
is the pion valence (sea) quark structure function and $G^N_{v(s)}(x_2)$
parameterizes the nuclear contribution.
Assuming isospin symmetry for $\pi^-$ we have

\begin{eqnarray}
& & f^\pi_v(x) = x{\overline u}(x) = xd(x), \nonumber \\
& & f^\pi_s(x) = x u_s(x) = x {\overline u}_s(x) = \dots = x{\overline
s}(x). \end{eqnarray}
The nuclear contributions can be similarly parameterized in terms
of valence and sea quark distributions of individual nucleons.
The normalization ($K$ factor) is measured to be $K\sim
1.75\pm 0.13$ while a perturbative analysis to first order in
$\alpha_s$ gives $K \sim 1.4$.
The structure functions are in principle functions of of $x_i$ and $Q^2$,
however, since we are describing the average data for
$36.0 < Q^2 < 72.3
\mbox{ GeV}^2$ we have suppressed the explicit $Q^2$
dependence. The meson structure functions
can equivalently be defined in terms of a diagonal matrix
element involving the commutator of two vector currents~\cite{IZ}

\begin{equation}
W^{\mu\nu}_i(x,Q^2) = {1\over {4\pi}}\int dz e^{iqz} \langle p
|[{\overline
\psi}_i\gamma^\mu \psi_i(z), {\overline \psi}_i\gamma^\nu \psi_i(0)] | p
\rangle
\label{wmunu}
\end{equation}
with $x = p\cdot q/M^2$, $Q^2 = -q^2 > 0$ and $i$ referring to a particular
flavor.
Since it is known that the total longitudinal momentum of the
hadron is only partially distributed among quarks, the
constituent quarks which are assumed to cary the entire momentum of
the hadron cannot be identified with the partons contributing to
deep inelastic structure functions.
 In the scaling
limit, $Q^2 \to \infty$, Eq.~(\ref{wmunu}) can be calculated using a
short distance expansion of the bilocal operator.
In perturbative QCD scaling violations can be described in terms
of a convolution of the partonic distributions defined at
a scale of reference $Q^2_0 < \infty$ and the
Altarelli-Parisi splitting functions
which characterize the single
parton response amplitudes for the change of scale due to
radiation of gluons~\cite{YAN}. Here we also use the convolution
 approach ~\cite{ALT} to relate our constituent quark model to the parton
model.
In a convolution model
the quark distribution function for a hadron $\alpha$, $q^\alpha_i$, is
represented by
 a product of the distribution function of a constituent quark,
$Q^\alpha_v$, in a hadron and the probability for a constituent quark to
fragment into a QCD parton $i$, $q_{i/v}$,

\begin{equation}
q^\alpha_i(x,Q^2) = \int_x^1 {{dy}\over y}Q^\alpha_v(x,Q^2_0)
q_{i/v}({x\over y},Q^2/Q^2_0).
\end{equation}
The constituent quark distributions are defined through
Eq.~(\ref{wmunu})
with the QCD fields replaced by an effective
constituent quark/gluon basis at $Q^2_0 \sim 1\mbox{ GeV}^2$.
The $Q^2$ evolution of $q_{i/v}(x/y,Q^2/Q^2_0)$  is governed by perturbative
QCD.
However for any value of $Q^2$
phenomenological input is still required.
For the average $Q^2 \sim 50\mbox{ GeV}$ of the Drell-Yan data the
valence and sea quark distributions are

\begin{eqnarray}
& & q^\pi_v(x) = \int_x^1 {{dy}\over y}Q^\pi_v(x) q_{v/v}({x\over y}),
\nonumber \\
& & q^\pi_s(x) = \int_x^2 {{dy}\over y}Q^\pi_v(x) q_{s/v}({x\over y}).
\label{qdef} \end{eqnarray}
For the valence quark contributions in $\pi^-$, $q_v = q_d=q_{\overline u}$,
while for the sea $q_s=q_d=q_{\overline d} = \cdots q_{\overline s}$.
The number and momentum sum rules are respectively

\begin{eqnarray}
& & \int_0^1 dx q^\pi_v(x) = 1, \nonumber \\
& & 2\int_0^1 dx x q^\pi_v(x) + 6\int_0^1 dx x q^\pi_s(x) = 1 - g_\pi,
\label{sum}
\end{eqnarray}
where $g_\pi$ represents the fraction of the gluon momentum in the pion
currently measured as $g_\pi \sim  0.47$~\cite{DY2}.
The parton distributions in a constituent quark, $q_{i/v}$, are usually
normalized to describe
the relevant features of the low-x hadron scattering phenomenology.
The Regge behaviour at small $x$ motivates
the following  parameterization~\cite{ALT}

\begin{eqnarray}
& & q_{v/v}(x) = {{ \Gamma\left(A + {1\over 2}\right) }
\over {\Gamma\left(1\over 2 \right)\Gamma(A) } }{1\over
\sqrt{x}}(1-x)^{A-1},\nonumber \\
& & q_{s/v}(x) = {C\over x}(1-x)^{D-1} \label{art}
\end{eqnarray}
with the parameters $A,D$ and $C$ constrained by the momentum sum
rule of Eq.~(\ref{sum}),

\[
{1\over {2A + 1}} + 6{C\over D} = 1 - g_\pi
\]
Each can be determined by comparing the calculated quark
distributions $q^\pi_v$ and $q^\pi_s$ to the data.
The matrix element which determines the constituent
quark distribution functions obtained from the light cone expansion of the
current product in Eq.~(\ref{wmunu}) has a very similar structure to
Eq.~(\ref{da}) that defines the quark distribution amplitude.
Using the techniques developed in the
 previous section the following expression for $Q^\alpha_v(x)$ is
 easily derived,

\begin{equation}
Q^\alpha_v(x) = \int {{d^2q^\perp_T}\over {(2\pi)^3}}J(q^\perp_T,\xi)
 |\Phi({\cal M})|^2
\left[1 + r_{PT}{{ {\cal M}^2 - (m_1+m_2)^2}\over
{\beta {\cal M}}}\right]^2.
\end{equation}

For $\alpha=\pi^-$, $m_1=m_2$ the integrand is a symmetric function of
$\xi =
2x - 1$ yielding $Q^{\pi^-}_{\overline u}(x) = Q^{\pi^-}_d(x)$. For the
kaon, $m_2=m_s
> m_1 = m_u = m_d$ and the nonstrange and strange quark distributions
correspond to $\xi = 2x - 1$ and $\xi = -2x + 1$ respectively.
Again ${\cal M} = {\cal M}(q^\perp_T,\xi)$ and $J(q^\perp_T,\xi)$ are
specified in the Appendix.

In Fig.~5 we plot valence quark structure functions for the pion
calculated in our model with
$r_{PT} = 1.0$ (solid line) obtained from the form factor fit
and the result without the tensor component having $r_{PT}=0$
 (dashed line) and compare with data. The comparison suggests
the importance
of the tensor term at large $x$, $0.6 < x < 0.9$ where the sensitivity to the
constituent quark distributions is the highest. In Fig.~6 we also compare
our results for the sea quark distributions (solid line) with the curve
used for experimental fitting (dashed line).
The value $C=0.086$ for the parameter in Eq.~(\ref{art}) is obtained
by requiring the theoretical and experimental curves to agree at $x=0$.
Fitting $A$ by the valence quark distribution yields $A=0.75$
and the sum rule of Eq.~(\ref{sum}) then gives
 $D=4.0$. These numbers are in a good agreement with the
ones obtained from the unpolarized nucleon structure function fits
confirming hadron independence of the splitting
functions $q_{i/v}(x)$~\cite{ALT,ZDAS}.
In Fig.~7 we also show our predictions for the strange (solid line) and
light quark structure functions (dashed line) in kaon.

\section{Summary and Conclusions}

Within the framework of our covariant variable front quark model,
we have generalized the constituent, $q{\overline q}$ pion wave function
and have studied the distribution amplitudes and structure functions.
The extended model provides excellent agreement with the experimental
data for the structure
functions and the electromagnetic form factor as well as a good
description
of the decay constant. The improved description is due to a tensor component
 in the wave function which is a
relativistic
correction in the rest frame.
  We have
computed the structure functions and the distribution amplitudes with
the wave function of Eq.~(\ref{wfnew}) in a noncovariant light cone
quantization scheme and find results are similar to the
 variable front model. This confirms front independence of our
results and
is consistent with a previous assertion ~\cite{AS} that different
front formulations do not lead to significant
differences in the predictions for various mesonic properties.
The magnitude of the tensor term which optimizes both the form factor and
structure function
 now indicate a camel shape for the distribution amplitude with the dip at
$\xi=0$ although not as profound as the one suggested by the old QCD sum
rules, but quite similar to recent nonlocal QCD sum rule calculations
and lattice results.
Since
the shape and the moments of the quark distribution amplitude are very
sensitive to the interference between the pseudoscalar and tensor
components of the wave function the detailed knowledge of former will
provide important information on the structure of the meson
wave function. As mentioned in the introduction, the distribution amplitude
is not directly measurable, however, information on the magnitude of the
lowest moments can  be extracted from the meson form factor
describing the hadronic part of the $e \mbox{ meson} \to e \gamma $
transitions~\cite{CZ,ASAW}. Finally, distribution amplitude studies provide
an effective forum for theoretical comparisons of alternative model
formulations which can provide significant insight into QCD dynamics.

\acknowledgments

 Financial support from U.S. D.O.E. grants
DE-FG05-88ER40461 and DE-FG05-90ER40589 is acknowledged.

\section{Appendix}

The variable change of Eq.~(\ref{qtoxi}) gives,

\[
q^3_T(q^\perp_T,\xi) = {{ \mu^2_1(q^\perp_T)(1-\xi)^2 -
\mu^2_2(q^\perp_T)(1+\xi)^2
} \over \sqrt{8(1-\xi^2)[\mu^2_1(q^\perp_T)(1-\xi) +
 \mu^2_2(q^\perp_T)(1+\xi)]} }
\]

where

\[
\mu^2_i(q^\perp_T) \equiv m^2_i + (q^\perp_T)^2,
\]

\begin{eqnarray}
& & k_{iL}(q^\perp_T,\xi) = \sqrt{\mu^2_i(q^\perp_T) +
(q^3_T(q^\perp_T,\xi))^2}, \nonumber \\
& & {\cal M}(q^\perp_T,\xi) = k_{1L}(q^\perp_T,\xi) + k_{2L}(q^\perp_T,\xi)
\nonumber \end{eqnarray}

and the Jacobian $J(q^\perp_T,\xi)$ is given by

\widetext

\begin{eqnarray}
& & J(q^\perp_T,\xi) = 2 {\xi \over {1-\xi^2}} [{\overline \mu}^2
 + (\Delta\mu)^2 - 2\xi{\overline \mu}\Delta\mu]
- {2\over {1-\xi^2}}\Delta\mu{\overline \mu}
- 2\xi{{ {\overline \mu}^2(\Delta\mu)^2 }
\over { {\overline \mu}^2 + (\Delta\mu)^2 - 2\xi{\overline \mu
}\Delta\mu }} \nonumber \\
& & + 2(1-\xi^2) {{ {\overline \mu}^3(\Delta\mu)^3 }
\over { {\overline \mu}^2 + (\Delta\mu)^2 - 2\xi{\overline \mu
}\Delta\mu }}, \nonumber
\end{eqnarray}

\narrowtext

with

\begin{eqnarray}
& & {\overline \mu} = {\overline \mu}(q^\perp_T) \equiv
{1\over 2}(\mu_1(q^\perp_T) + \mu_2(q^\perp_T)), \nonumber \\
& & \Delta\mu = \Delta\mu(q^\perp_T) \equiv
{1\over 2}(\mu_1(q^\perp_T) - \mu_2(q^\perp_T)). \nonumber
\end{eqnarray}

\newpage

\figure{FIG. 1a,b. Pion electromagnetic form factor. Solid line shows the
result obtained with the $q{\overline q}$ wave function of Eq.~(2.12)
for $r_{PT} = 1.0$. Dashed line is the result of the covariant model
of Ref.~[4], dash-doted and doted lines are the results of
relativistic models from Refs.~[2] and~[3]
respectively. Data is taken from Ref.~[21].}

\figure{FIG. 2. Pion distribution amplitude with ($r_{PT}=1$, solid curve)
and without ($r_{PT} = 0$, dashed curve) the
contribution from the tensor term in the $q{\overline q}$ wave function.}
 
\figure{FIG. 3. Same as Fig.2 for the $K$ distribution amplitude.}

\figure{FIG. 4. $r_{PT}$ dependence of the pion distribution amplitude.

\figure{FIG. 5. Valence quark pion structure function $f^\pi_v(x) =
xq_v(x)$ (curve convention same as in FIG. 2.). Data is taken
from Ref.~[11]. }
\figure{FIG. 6. Sea quark pion structure function. Solid line gives our
model prediction with $r_{PT}=1.$, dashed line shows experimental
parameterization, $f^\pi_s(x) = 0.173(1-x)^{8.4}$ of Ref.~[11].}

\figure{FIG. 7. Valence light quark (dashed line) and strange quark
(solid line) kaon structure functions.}
\newpage

\begin{table}
\caption{$\pi$ and $K$ mesons decay constant calculated with the formula
of Eq.~(3.3). The two set of results correspond to the
use of spin averaged, constituent meson masses $M$ and dynamical masses
${\cal
M}$ respectively, ($m_u=m_d=250\mbox{MeV}$, $m_s = 480\mbox{MeV}$, $\beta =
250 \mbox{MeV}$, $r_{PT} = 1.0$) }
\begin{tabular}{|r|c|c|}
\hline
 & $f_\pi[\mbox{MeV}]$ & $f_K[\mbox{MeV}]$ \\
$M$ & 101. & 137.  \\
\hline
${\cal M}$ & 75. & 104. \\
\hline
exp. & 93.2 & 113. \\
\hline
\end{tabular}
\end{table}

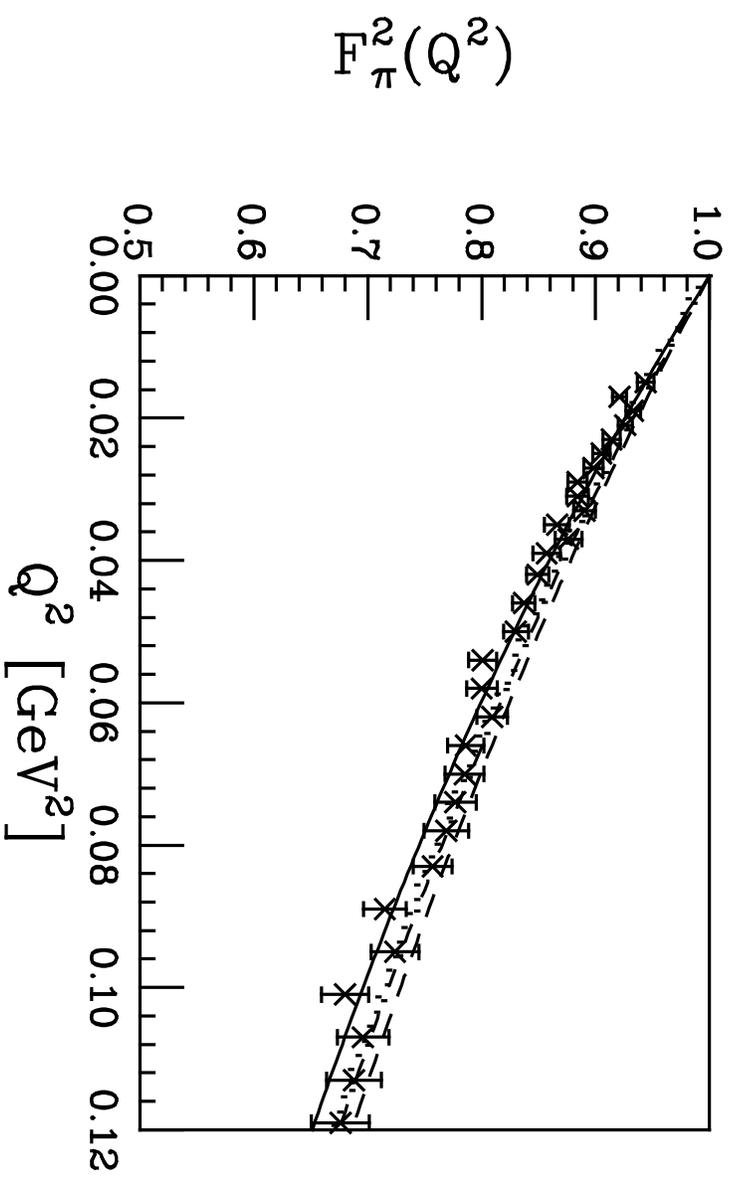
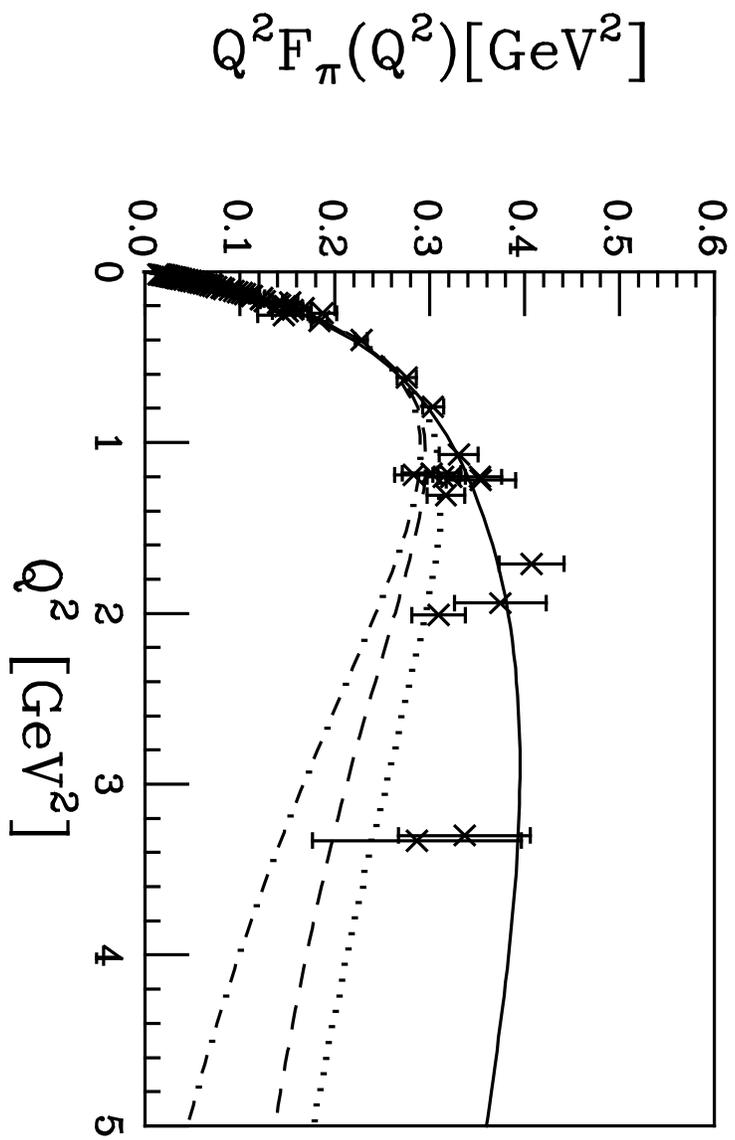
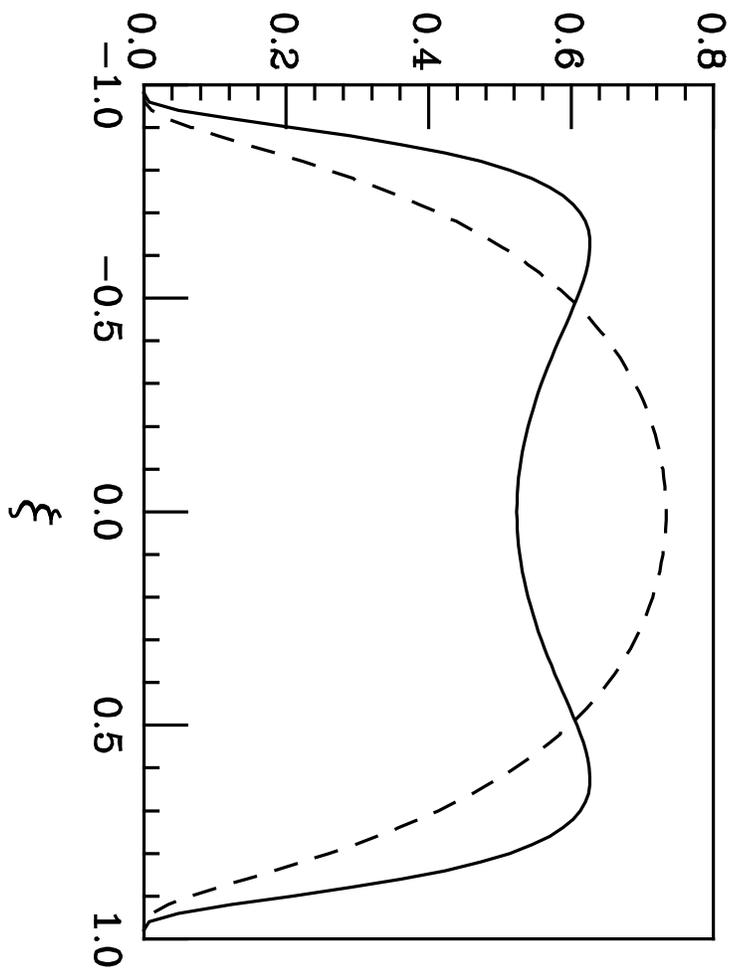
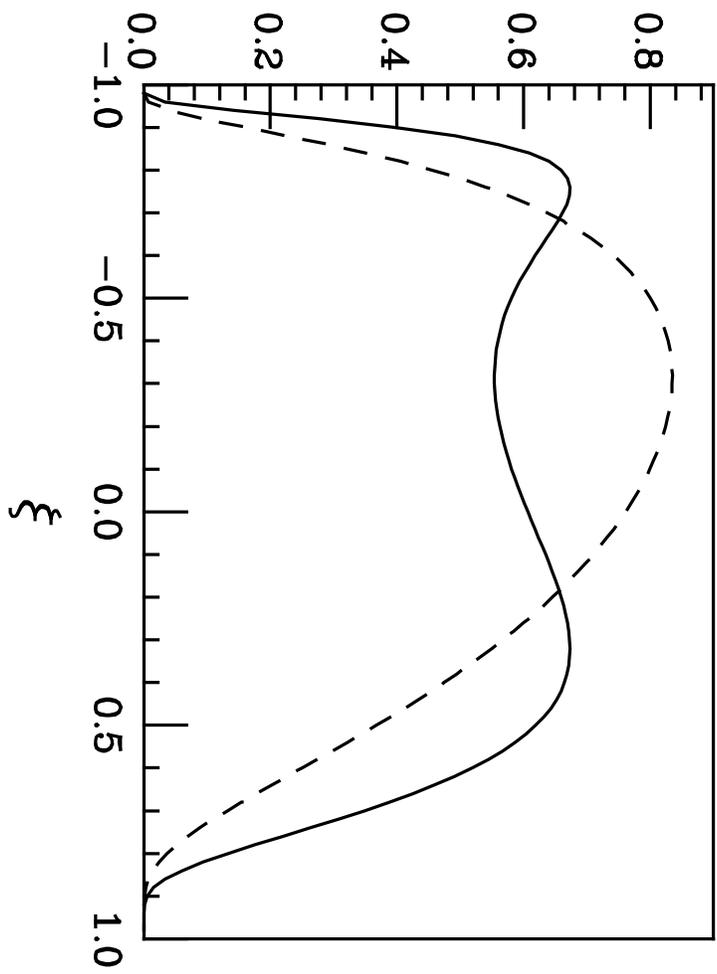
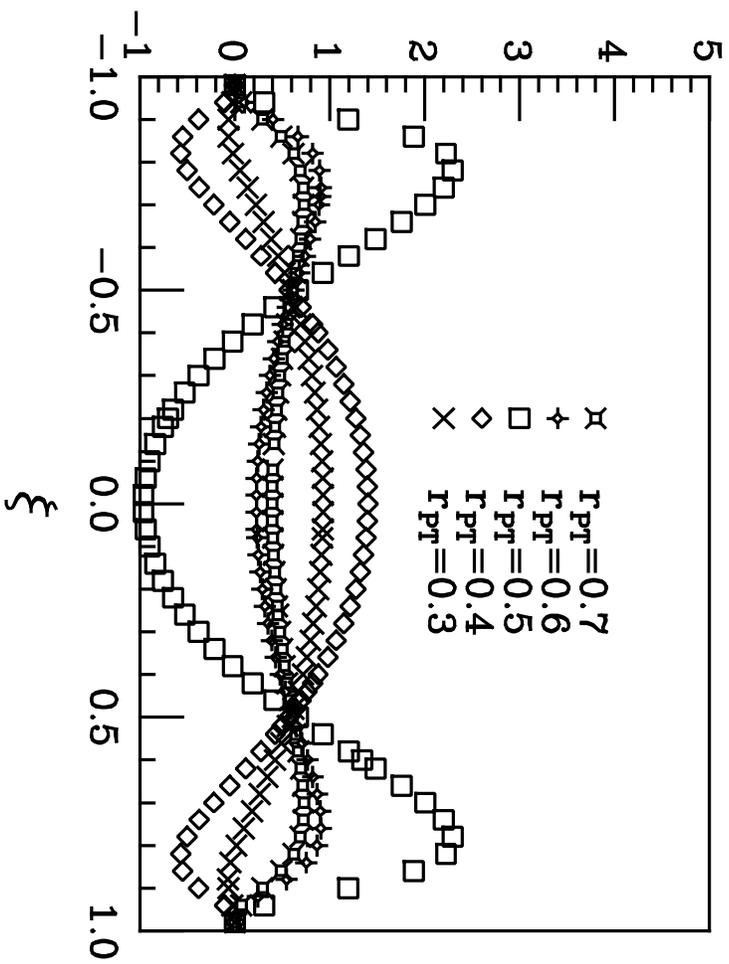
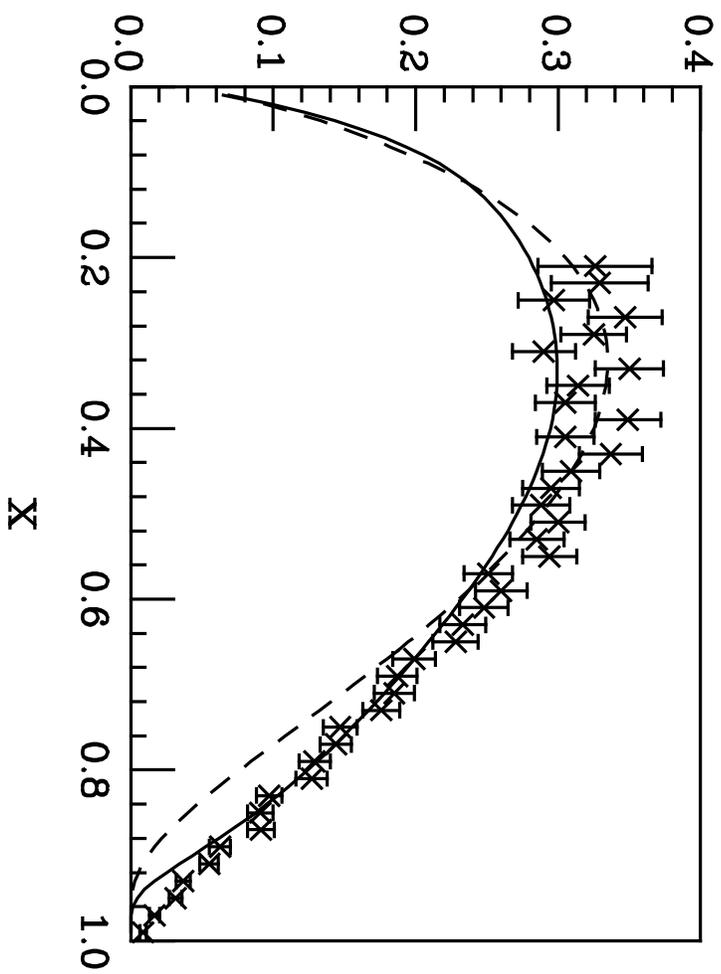
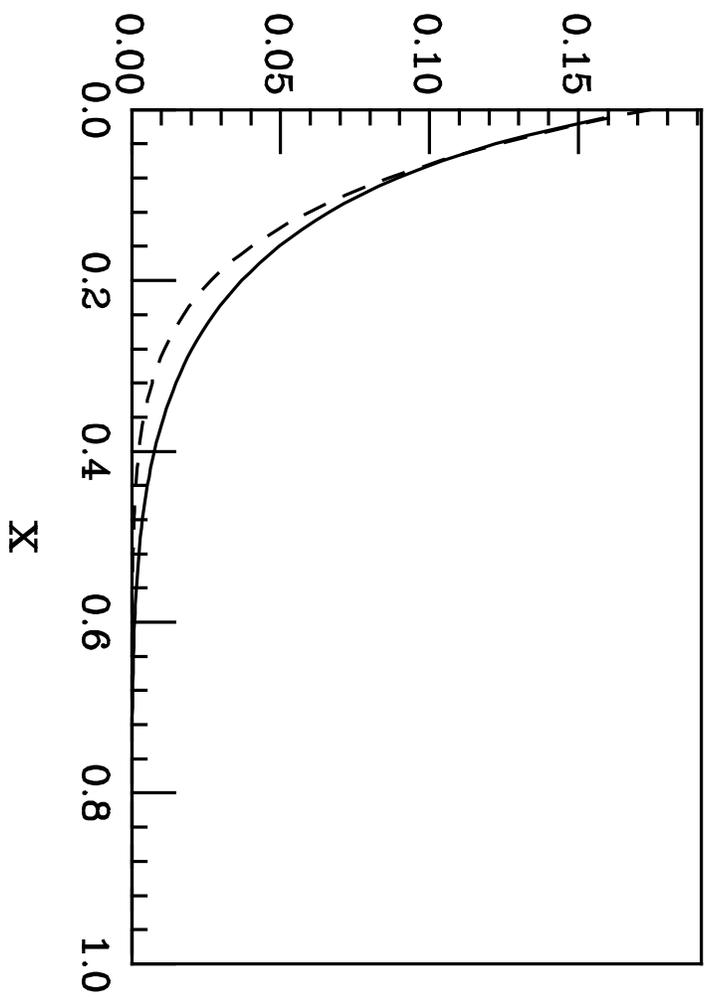
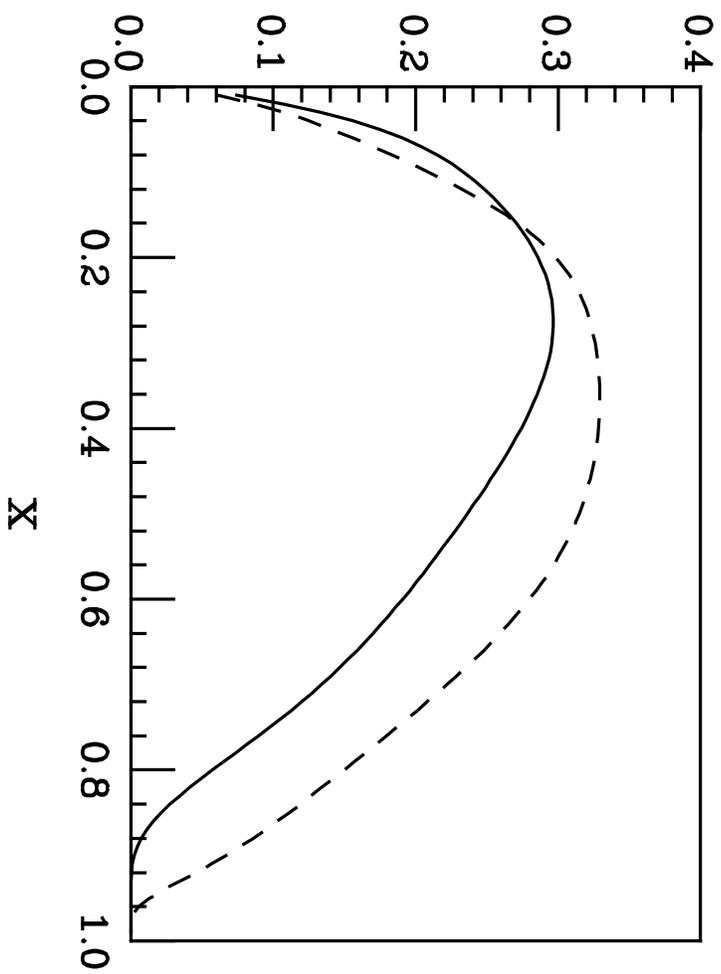
\end{document}